# Self-supervised Deep Unrolled Reconstruction Using Regularization by Denoising

Peizhou Huang, Chaoyi Zhang, Xiaoliang Zhang, Xiaojuan Li, Liang Dong, Leslie Ying

***Abstract*—Deep learning methods have been successfully used in various computer vision tasks. Inspired by that success, deep learning has been explored in magnetic resonance imaging (MRI) reconstruction. In particular, integrating deep learning and model-based optimization methods has shown considerable advantages. However, a large amount of labeled training data is typically needed for high reconstruction quality, which is challenging for some MRI applications. In this paper, we propose a novel reconstruction method, named DURED-Net, that enables interpretable self-supervised learning for MR image reconstruction by combining a self-supervised denoising network and a plug-and-play method. We aim to boost the reconstruction performance of Noise2Noise in MR reconstruction by adding an explicit prior that utilizes imaging physics. Specifically, the leverage of a denoising network for MRI reconstruction is achieved using Regularization by Denoising (RED). Experiment results demonstrate that the proposed method requires a reduced amount of training data to achieve high reconstruction quality among the state-of-art of MR reconstruction utilizing the Noise2Noise method.***

***Index Terms*—Magnetic resonance image reconstruction, deep neural network, self-supervised, regularization by denoising.**

## I. INTRODUCTION

MAGNETIC resonance imaging (MRI) is one of the premier imaging modalities used clinically due to its ability to provide anatomic details and superior soft-tissue contrast. However, a major challenge of MRI is its long acquisition time. Many techniques have been developed to increase the MRI speed by acquiring the k-space data below the Nyquist rate, such as parallel imaging [1, 2]. Over the past decades, compressed sensing (CS) [3] has become an important technique to accelerate MRI. Many optimization methods with different regularization models have been developed for CS-MRI, such as learned dictionaries [4, 5], low-rankness [6, 7], and manifold models [8, 9], leading to robust and practical algorithms with excellent performance.

In the past few years, deep learning [10] techniques provide excellent end-to-end learning capabilities in the field of computer vision, ranging from denoising, deblurring, to super-resolution and image restoration. Inspired by that success, deep neural networks have been applied to MRI reconstruction from reduced measurements [11-17]. Different from CS-MRI methods which solve the ill-posed inverse problem with mathematical constraints using optimization algorithms, these deep learning-based methods use a large amount of labeled training data to learn the complicated relationship between the aliased image (inverse Fourier transform of undersampled k-space) and the ground truth image (inverse Fourier transform of fully-sampled k-space), which is represented with standard convolutional neural networks (CNN). These deep learning-based methods are able to address the limitations that the CS-MRI methods typically need sophisticated priors with hand-tuned parameters and iterative algorithms with long runtime. However, it can be challenging to obtain a large amount of labeled training data in certain applications such as dynamic imaging, relaxometry imaging, and diffusion imaging.

In the meanwhile, several studies have applied deep learning to model-based optimization methods in MR image reconstruction by unrolling the iterative optimization to a deep network [18-20]. For example, Yang et al [18] introduced a single-coil MR reconstruction method by unrolling the alternating direction method of multipliers (ADMM) algorithm to several self-defined layers to learn the regularization parameters in ADMM. Hammernik et al [19] unrolled the variational models for iterative reconstruction to a deep network to learn the regularization parameters, as well as the nonlinear operators. Zhang et al [20] unrolled the iterations in the ISTA algorithm to layers of networks to learn the image transformation and parameters. These unrolling-based methods usually demand less training data due to the physics-driven architecture and the reduced number of training parameters compared with the standard networks.

More recently, several methods have been proposed to incorporate data consistency as additional information in the deep learning-based reconstruction methods. Those methods leverage the benefits of both the physical model and standard network structures. They took advantage of CNN's powerful capability of learning the prior model using a deep architecture and integrated it with the model-based inverse problem [21-24]. For example, Aggarwal et al. [21] combined a CNN denoiser with the conjugate gradient optimization scheme for MRI

This work was supported in part by the NIH/NIAMS R01 AR077452 and NIH U01 EB023829.
P. Huang and X. Zhang are with the Biomedical Engineering Department, State University of New York at Buffalo, Buffalo, NY 14260 USA.
C. Zhang is with Electrical Engineering Department, State University of New York at Buffalo, Buffalo, NY 14260 USA.
X. Li is with Program of Advanced Musculoskeletal Imaging (PAMI), Cleveland Clinic, Cleveland, Ohio 44103, USA
D. Liang is with Paul C. Lauterbur Research Center for Biomedical Imaging, SIAT, CAS, Shenzhen 518055, P.R.China
L. Ying is with the Biomedical Engineering Department and Electrical Engineering Department, State University of New York at Buffalo, Buffalo, NY 14260 USA.

reconstruction. Schlemper et al. [22] created a cascade architecture that alternates between an intermediate de-aliasing CNN and data consistency for dynamic MRI reconstruction. Eo et al [23] applied the image domain CNN and k-space CNN with interleaved data consistency in MRI reconstruction that allows outstanding detail restoration. Cheng et al [24] proposed to embed both data consistency and prior information in the network such that both can be learned jointly. A more comprehensive summary of deep learning-based reconstruction algorithms can be found in [25].

However, the above-mentioned deep learning-based reconstruction methods require ground truth images as the labeled training data. Some methods have been developed to use the undersampled data only for self-supervised learning [26-35]. Among these methods, some generate paired training data from the undersampled data [26-28], some use the Generative Adversarial Networks (GAN) to guide the training process [29-32], and others use statistical properties of the data to reconstruct the image [33-35]. Among them, SSDU [26] splits the acquired k-space data into two disjoint sets, one used as the input and the other as the target for cross-validation, so that self-supervised learning is performed without fully sampled data. In addition, Noise2Noise [36], a CNN-based denoiser, has been used to reconstruct clean images from aliased images without ground-truth training images. Built upon the success of plug-and-play methods that solve a regularized inverse problem by sequentially applying image denoising with high-performing denoisers, RARE [37] uses a pre-trained Artifacts2Artifacts network (equivalent to Noise2Noise in [36]) as the denoiser to perform MR image reconstruction with self-supervised learning. However, the limitation of RARE is that the image reconstruction quality heavily relies on the pre-trained CNN denoiser whose training requires a large training size. As a result, the RARE reconstruction quality is not optimal.

In this paper, we focus on the line of work based on Noise2Noise [36]. We propose a novel deep unrolled self-supervised method powered by regularization by denoising (RED) [38] for MRI reconstruction, named DURED-Net. The method integrates Noise2Noise with RED. Unlike the existing Noise2Noise-based methods for MR reconstruction, RARE, where the denoising network is pre-trained with self-supervised learning and then plugged into the iterative procedures of RED directly, the denoising engine in the proposed method is trained along with the unrolled ADMM iterations of RED so that the network parameter depends not only on the training samples but also on the imaging model of the inverse problem. Such integrated training makes self-supervised learning to be more efficient for the specific imaging model. In addition, the ADMM iteration is also treated as an unrolled network whose parameter is learned using the training data. In particular, although RARE also combines the Noise2Noise network with RED, our proposed method is superior in that it not only leverages the capability of Noise2Noise but also unrolls the ADMM-based iterations of RED as another network and learns the parameters of both networks jointly during the training process. As a result, the proposed method demonstrates improved reconstruction quality, alleviated overfitting issues, lessened requirements on training size, and reduced sensitivity to network hyperparameters (e.g., layers, etc.).

This paper is organized as follows. Section II presents the background of Noise2Noise, a self-supervised method for image denoising, and RED. Section III presents the proposed unrolled self-supervised method and illustrates each module of the network architecture. Section IV describes the datasets used for evaluation and the implementation details of the proposed model. In section V, we present promising results compared to other supervised and self-supervised methods. The conclusion is presented in Section VI.

## II. BACKGROUND

### A. Regularizing image reconstruction with denoising

In the context of compressed sensing, the purpose is to reconstruct the latent image $\mathbf{x}$ from its undersampled measurement $\mathbf{y} = \mathbf{A}\mathbf{x}$, where $\mathbf{A}$ is a measurement matrix. In the single-coil acquisition, we have $\mathbf{A} = \mathbf{SF}$ where $\mathbf{S}$ is the sampling pattern, and $\mathbf{F}$ is the Fourier transform. And in the multi-coil acquisition, $\mathbf{A} = \mathbf{SFC}$ where $\mathbf{C}$ includes coil sensitivity information. Since the problem is ill-posed, regularization needs to be adopted to reconstruct the latent image. Mathematically, this entails solving an optimization problem by minimizing this objective function:

$$\hat{\mathbf{x}} = \arg\min_{x} \frac{1}{2}\|\mathbf{A}\mathbf{x} - \mathbf{y}\|^2 + \lambda \rho(\mathbf{x}). \quad (1)$$

where $\frac{1}{2}\|\mathbf{A}\mathbf{x} - \mathbf{y}\|^2$ represents data consistency, $\rho(\mathbf{x})$ the regularization function, and $\lambda$ the regularization parameter.

As many denoising algorithms have demonstrated their success in the past years, several methods have been proposed to leverage denoising algorithms in image reconstruction [38-39]. Among these works, Regularization by Denoising (RED) [38] demonstrates its advantages of flexibility in using the denoising engine for regularization in inverse problems. RED constructs an explicit formulation as the regularization function with an arbitrary image denoiser:

$$\hat{\mathbf{x}} = \arg\min_{\mathbf{x}} \frac{1}{2}\|\mathbf{A}\mathbf{x} - \mathbf{y}\|^2 + \frac{1}{2}\lambda \mathbf{x}^T(\mathbf{x} - f(\mathbf{x})). \quad (2)$$

where $f(\cdot)$ is a denoiser engine. There are two important properties in RED: (1) The gradient of the prior is given by $\nabla \rho(\mathbf{x}) = \mathbf{x} - f(\mathbf{x})$, which avoids differentiating the denoisers function. (2) $\rho(\mathbf{x})$ is a convex functional. With the development of plug-and-play methods, the denoiser engine can be plugged into an iterative scheme in different ways. All those schemes involve a data consistency term-related subproblem and a regularization term that contains the denoiser engine related to the denoising subproblem. Consequently, many highly effective denoising algorithms can be used to solve reconstruction problems.

Recently, with the powerful denoising ability of the neural network, we can see that the neural network can be as a denoiser prior and plugged into an iterative scheme in different ways [21, 37]. However, most of these existing works relied on a pre-trained network where the network parameters are pre-learned and not changed along with the iterative procedure of the corresponding algorithm.

### B. Deep learning reconstruction using unrolling

Deep learning has also been used in MRI reconstruction recently to learn the mapping between the aliased image $\hat{\mathbf{x}}_i$ from undersampled k-space data (input) and the ground truth image from fully sampled data (target) $\mathbf{x}_{GT}$ using a large number of training pairs $(\hat{\mathbf{x}}_i, \mathbf{x}_{GT})$. However, deep learning methods cannot guarantee data consistency nor utilize sophisticated priors for improved performance. Therefore, there have been several works that unroll the iterative procedures in MRI image reconstruction to a deep network with a certain architecture so that the regularization functions and parameters that used to be fixed become learnable using the training data, such as ADMM-net [18], variational-net [19], ISTA-net [20], and MoDL [21].

These unrolling-based methods have the advantage of utilizing the MR physical model while embracing the benefit of a deep network. It has been shown that decoupling the data consistency term and regularization term allows a wide variety of existing inverse algorithms to be combined with a neural network. The unrolling-based method commonly starts from a regularized optimization algorithm for Eq. (1) but unrolls iterations of the algorithm to stages of the neural network. The method overcomes the issue of selecting regularization functions and parameters. For example, ADMM-Nets unroll the ADMM iterations to a deep network from the following minimization problem

$$\hat{\mathbf{x}} = \arg\min_{\mathbf{x}} \frac{1}{2}\|\mathbf{A}\mathbf{x}-\mathbf{y}\|^2 + \sum_{l=1}^{L}\lambda_l\, g(\mathbf{D}_l\mathbf{x}). \quad (3)$$

It learns the nonlinear transforms $\mathbf{D}_l$, regularization function $g(\cdot)$, and parameters $\lambda_l$ using training data.

### C. Noise2Noise

Deep learning has shown success in image denoising [36, 40]. In supervised learning, the network learns the relationship between the noisy image (input) $\hat{\mathbf{y}}_1^i$ and its clean counterpart (target) $\mathbf{y}_{GT}^i$ using a large number of training pairs $(\hat{\mathbf{y}}_1^i, \mathbf{y}_{GT}^i)$ by minimizing the mean squared error (L2):

$$\arg\min_{\theta} \frac{1}{n}\sum_{i=1}^{n}\|f(\hat{\mathbf{y}}_1^i, \theta) - \mathbf{y}_{GT}^i\|^2. \quad (4)$$

where $f(\hat{\mathbf{y}}_1^i, \theta)$ is the denoising neural network with network parameter $\theta$, $n$ is the total number of training samples. Because the clean image is not always available, Noise2Noise has been proposed as a deep learning method without the need for the ground-truth clean image in training. It replaces the noisy-clean training pairs with noisy-noisy ones when training the network:

$$\arg\min_{\theta} \frac{1}{n}\sum_{i=1}^{n}\|f(\hat{\mathbf{y}}_1^i, \theta) - \hat{\mathbf{y}}_2^i\|^2. \quad (5)$$

where $\hat{\mathbf{y}}_1^i$ and $\hat{\mathbf{y}}_2^i$ are both noisy images of the same clean one. It was demonstrated in [36] that such training is equivalent to training with the ground truth images when the following conditions are satisfied:

1. The number of training samples $n$ is infinite;
2. The conditioned expectation $\mathbb{E}\{\hat{\mathbf{y}}_2^i|\hat{\mathbf{y}}_1^i\} = \mathbf{y}_{GT}^i$;
3. $\hat{\mathbf{y}}_1^i$ and $\hat{\mathbf{y}}_2^i$ are from the same image corrupted by independent noise of the same distribution.

That is, minimizing the loss function of Eq. (5) is the same as that of Eq. (4) when the above conditions are satisfied.

For MRI acquisition, $\hat{\mathbf{y}}_1^i$ and $\hat{\mathbf{y}}_2^i$ can be acquired in two separate scans, but it is not efficient. In practice, we can acquire $\hat{\mathbf{y}}_1^i$ and $\hat{\mathbf{y}}_2^i$ simultaneously in a single scan. Specifically, the two sampling patterns for $\hat{\mathbf{y}}_1^i$ and $\hat{\mathbf{y}}_2^i$ are first generated independently based on the above-mentioned conditions for Noise2Noise. The details of the sampling design are provided in Section III.C. The two patterns, which are not necessarily disjoint, are then combined (taking the union) into one pattern, which is used to acquire the k-space data $\hat{\mathbf{y}}_{1+2}^i$ in a single scan. After the scan, the acquired data are separated into $\hat{\mathbf{y}}_1^i$ and $\hat{\mathbf{y}}_2^i$ using the two prior generated patterns, where the overlapped data go to both $\hat{\mathbf{y}}_1^i$ and $\hat{\mathbf{y}}_2^i$. In this way, we can obtain the training data $\hat{\mathbf{y}}_1^i$ and $\hat{\mathbf{y}}_2^i$ within a single scan, while the scan is still accelerated with undersampling.

## III. PROPOSED METHOD

### A. Problem formulation

In RED, the denoiser $f(\cdot)$ is plugged into the optimization procedure of Eq. (1), and the splitting variable method is adopted to decouple the data consistency term and a regularization term. Using ADMM [41], Eq. (1) is first reformulated as a constrained optimization problem by introducing an auxiliary variable $v$:

$$(\hat{\mathbf{x}}, \hat{\mathbf{v}}) = \arg\min_{\mathbf{x},\mathbf{v}} \frac{1}{2}\|\mathbf{A}\mathbf{x}-\mathbf{y}\|^2 + \lambda\frac{1}{2}\mathbf{v}^T(\mathbf{v}-f(\mathbf{v}))\ s.t.\ \mathbf{x}=\mathbf{v}. \quad (6)$$

Then solving Eq. (6) is equivalent to minimizing the augmented Lagrangian function:

$$\mathcal{L}(\mathbf{x},\mathbf{v},\mathbf{u}) = \frac{1}{2}\|\mathbf{A}\mathbf{x}-\mathbf{y}\|^2 + \lambda\frac{1}{2}\mathbf{v}^T(\mathbf{v}-f(\mathbf{v})) + u^T(\mathbf{x}-\mathbf{v}) + \frac{\beta}{2}\|\mathbf{x}-\mathbf{v}\|^2. \quad (7)$$

Finally, finding the minimum for $\mathcal{L}$ can be solved by a sequence of subproblems in the form

$$\begin{cases} \mathbf{x}^n = \arg\min_{\mathbf{x}} \left\{\frac{1}{2}\|\mathbf{A}\mathbf{x}-\mathbf{y}\|_2^2 + \frac{\beta}{2}\|\mathbf{x}-\tilde{\mathbf{x}}^{n-1}\|_2^2\right\} \\ \mathbf{v}^n = \arg\min_{\mathbf{v}} \left\{\frac{\beta}{2}\|\mathbf{v}-\tilde{\mathbf{v}}^{n-1}\|_2^2 + \frac{\lambda}{2}\mathbf{v}^T(\mathbf{v}-f(\mathbf{v}^{n-1}))\right\} \\ \mathbf{u}^n = \mathbf{u}^{n-1} + (\mathbf{x}^n - \mathbf{v}^n), \end{cases} \quad (8)$$

where $n$ refers to the iteration index in ADMM, $u^n$ is the scaled Lagrange multiplier, $\tilde{\mathbf{x}}^{n-1} = \mathbf{v}^{n-1} - \mathbf{u}^{n-1}$ and $\tilde{\mathbf{v}}^{n-1} = \mathbf{x}^n + \mathbf{u}^{n-1}$ are the intermediate variables.

*1) Solving subproblem $\mathbf{x}^n$ in Eq. (8)*: The subproblem of $x^n$ can be solved analytically by:

$$\mathbf{x}^n = (A^H\mathbf{A} + \beta\mathbf{I})^{-1}(A^H(\mathbf{y}) + \beta(\mathbf{v}^{n-1}-\mathbf{u}^{n-1})). \quad (9)$$

Here we discuss the single-coil and multi-coil cases separately. In the single-coil reconstruction, we take the linear combination between $\mathbf{v}^{n-1} - \mathbf{u}^{n-1}$ and the original sampled k-space measurement to perform the data consistency in the reconstruction layer $\mathbf{x}^n$ [22]. Therefore, $\mathbf{x}^n$ has a simple closed-form solution

$$\mathbf{x}^n = \mathbf{F}^{-1}(\tilde{\mathbf{x}}^n) \quad (10)$$

$$\tilde{\mathbf{x}}^n(k) = \begin{cases} \dfrac{\tilde{\mathbf{y}}(k)+\beta\mathbf{y}(k)}{1+\beta} & \text{if } k \text{ is sampled} \\ \tilde{\mathbf{y}}(k) & \text{otherwise} \end{cases} \quad (11)$$

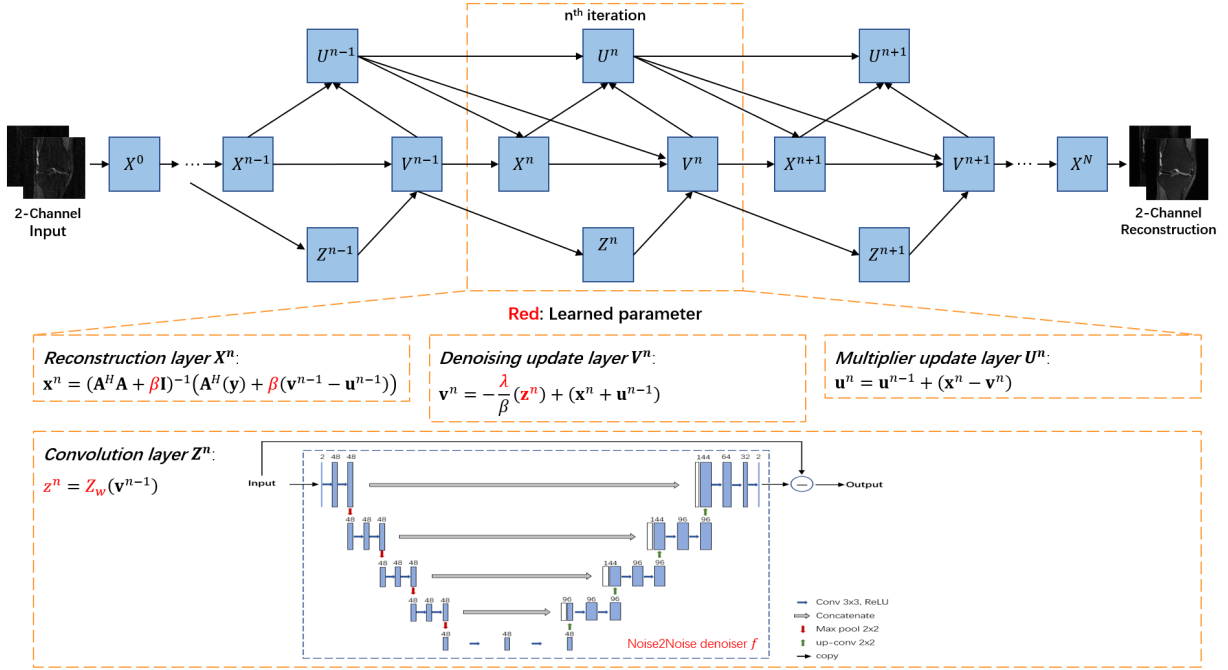

Figure 1. The proposed deep learning framework for MRI image reconstruction. The network structure is defined over the iterative procedures of Eq. (16). The four procedures correspond to four layers, which are named as reconstruction layer $X^n$, Convolution layer $Z^n$, Denoising update layer $V^n$, and Multiplier update layer $U^n$. The variables in red are learned during training. The graph illustrates the unrolled architecture with 3 modules in one training epoch.

where $\mathbf{y}$ is the acquired data and $\tilde{\mathbf{y}} = \mathbf{F}(\mathbf{v}^{n-1} - \mathbf{u}^{n-1})$. In the multi-coil reconstruction, considering the difficulty in finding the inverse of the operator $\mathbf{A}^H\mathbf{A} + \beta\mathbf{I}$, we use a few conjugate gradient iterations to approximate the solution in Eq. (9).

*2) Solving subproblem $\mathbf{v}^n$ in Eq. (8)*: In RED, it was shown that for any image denoiser, the gradient of RED regularization obeys the gradient rule that

$$\nabla \rho(\mathbf{x}) = \mathbf{x} - f(\mathbf{x}). \quad (12)$$

Then, any minimization of $v^n$ in (8) must yield $\nabla v^n = 0$, so that the gradient of $v^n$ can be represented in the form of

$$\lambda(\mathbf{v}^n - f(\mathbf{v}^{n-1})) + \beta(\mathbf{v}^n - \mathbf{x}^n - \mathbf{u}^{n-1}) = 0. \quad (13)$$

Interestingly, Eq. (13) implies that the minimization of $v^n$ contains the denoising residual from $\mathbf{v}^{n-1}$. Therefore, in this paper, a CNN is constructed to learn such deep denoising residual in the form of

$$\mathbf{z}^n = Z_w(\mathbf{v}^{n-1}) = \mathbf{v}^n - f(\mathbf{v}^{n-1}). \quad (14)$$

where $Z_w$ denotes the CNN with the learned network parameter $w$. Therefore, CNN can be learned along with solving subproblems $\mathbf{v}^n$ in ADMM by taking the input $\mathbf{v}^{n-1}$, and the solution of $\mathbf{v}^n$ in (8) can be represented directly as

$$\mathbf{v}^n = -\frac{\lambda}{\beta}(\mathbf{z}^n) + (\mathbf{x}^n + \mathbf{u}^{n-1}). \quad (15)$$

*3) Updating $\mathbf{u}^n$ using Eq. (8).*

Combining the above three subproblems, the solution to Eq. (8) at the $n$-th iteration is updated as follows:

$$\begin{cases} X^n: \mathbf{x}^n = (\mathbf{A}^H\mathbf{A} + \beta\mathbf{I})^{-1}(\mathbf{A}^H(\mathbf{y}) + \beta(\mathbf{v}^{n-1} - \mathbf{u}^{n-1})) \\ Z^n: \mathbf{z}^n = Z_w(\mathbf{v}^{n-1}) \\ V^n: \mathbf{v}^n = -\frac{\lambda}{\beta}(\mathbf{z}^n) + (\mathbf{x}^n + \mathbf{u}^{n-1}) \\ U^n: \mathbf{u}^n = \mathbf{u}^{n-1} + (\mathbf{x}^n - \mathbf{v}^n). \end{cases} \quad (16)$$

### B. Network architecture

We unroll the iterations in Eq. (16) to a deep neural network such that all the network parameters can be learned through training instead of being selected empirically. The architecture of the deep network is shown in Fig. 1. The four steps in Eq. (16) are represented as four layers in the neural network including the reconstruction layer $X^n$, the convolution layer $Z^n$, the denoising update layer $V^n$, and the multiplier update layer $U^n$. The $n$-th iteration corresponds to the $n$-th module of the deep network enclosed with the dashed box. Figure 1 shows an example with three iterations in one training epoch. All parameters $\{\lambda, \beta\}$ and the denoiser $f$ that are predetermined in the original RED-ADMM algorithm are now learned through training in the proposed method.

Next, we provide details for each module of the proposed DURED-Net.

*1) Reconstruction layer $X^n$*: This layer reconstructs an image of the current iteration or layer under the given $\{\mathbf{v}^{n-1}, \mathbf{u}^{n-1}\}$ according to the first item of Eq. (16), $X^n$: $\mathbf{x}^n = (\mathbf{A}^H\mathbf{A} + \beta\mathbf{I})^{-1}(\mathbf{A}^H(\mathbf{y}) + \beta(\mathbf{v}^{n-1} - \mathbf{u}^{n-1}))$. In the input layer $x^0$, $v^0$ are initialized in two channels representing the real and imaginary parts of the zero-filled Fourier reconstruction, and $\mathbf{u}^0$ is initialized to zero. For $n \geq 1$, the input and output of $\mathbf{x}^n$ are both complex-valued data. The hyper-parameter $\beta$ is set as a learnable parameter, which is initialized to 10.

*2) Convolution layer $Z^n$*: This layer performs a convolution operation to find the deep denoising residual of $\mathbf{v}^{n-1}$ according to the second term of Eq. (16), $Z^n: \mathbf{z}^n = Z_w(\mathbf{v}^{n-1})$. The network adopts a subtract structure between the input $\mathbf{v}^{n-1}$ and the denoiser network. Therefore, the denoising network $f_\theta$ is learned implicitly inside $Z_w$. We used a U-net [42] structure in the convolution layer $Z^1$ in our DURED-Net as shown in Fig. 1, the same as that used in Noise2Noise. Because standard CNNs processes real-valued data, both the input and output of $Z_w$ have two channels representing the real and imaginary parts of the complex data.

*3) Denoising update layer $V^n$*: This layer is to solve the denoising constrained problem to the actual minimization with the deep denoising residual $z^n$ with known $\{x^n, u^{n-1}\}$ according to the third item $V^n$ in Eq. (16). The two-channel output of the convolution layer $Z^n$ is combined into a single-channel input of complex value and used as the input of $V^n$. The hyper-parameter $\lambda$ is set as a learnable parameter, which is initialized to 10.

*4) Multiplier update layer $U^n$*: This layer is used to update the Lagrange multiplier $u^n$ under the last term $U^n$ in Eq. (16).

In the final reconstruction layer $X^n$, the complex output image is decomposed into the real and imaginary channels to calculate the $L_2$ loss.

Considering the increased memory and computational costs with more layers, here we choose a two-module network as shown in Fig. 2a, where the initial variable $X^0$ and $V^0$ are zero-filled input images from the undersampled data, $U^0$ is initialized to zeros. Networks with more modules are discussed in Section V. Figure 2b shows the intermediate output of each layer, where $\{X^i\}_{i=1}^2$ is the updated reconstruction, $V^1$ is the updated image in the regularization, $Z^1$ is the deep learned residual structure that is used for updating $V^1$, and $U^1$ is the updated dual variable.

It is worth noting that the input and output of the network, which are the zero-filled images from randomly undersampled data with different sampling patterns, meet conditions #2 and #3 of Noise2Noise for self-supervised training. The details of how to meet the conditions are elaborated further in Section III.C. As a result, similar to Noise2Noise, the proposed network can also generate a desired Nyquist-sampled image while using unlabeled undersampled training data only for training. This is significant in many dynamic imaging applications where the Nyquist-sampled data are impossible to acquire due to motion. The key benefit of the proposed scheme over [37] is that the denoising layer is not pre-trained but is learned jointly with other layers with data consistency constraints from the forward imaging model. The denoising layer is updated using the training data in each epoch so that the entire network performs better than the one with a pre-trained denoiser.

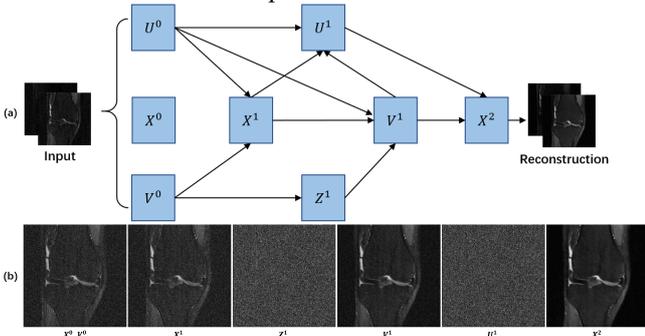

Figure 2. Simplified network. (a) Network architecture with two modules. (b) Intermediate outputs of layers in (a).

### C. Design of sampling pattern

Recall in Section I.C, there are three conditions for the self-supervised training in Noise2Noise to perform well [36]. In our context of image reconstruction, condition #2 means that if we take an infinite number of aliased images that are from independently generated random samples and average them (to approximate the expectation), then the average should be equal to the aliasing-free, ground truth image. Condition #3 means the k-space of an image is undersampled by two different sampling patterns that are randomly generated with the same probability density function. The second condition requires that the expectation of the random undersampling k-space spectrum be equal to the full sampling of the spectrum. To meet the requirements, Noise2Noise used a Bernoulli sampling where each k-space location has a probability $p(k) = e^{-\lambda |k|}$ of being selected for acquisition, where $k$ is the k-space location, and the sampled values are further weighted by $\frac{1}{p(k)}$. Here, we propose an improved random sampling pattern, where the probability density function (PDF) is

$$p(k) = e^{\left(\frac{-1}{\mu}|k|\right)^\alpha} \quad (17)$$

with $\mu$ and $\alpha$ the parameters to control the sampling location. The sampled values are also weighted by the inverse of the PDF. The proposed sampling pattern can be used for 1D, 2D, and even higher dimensional undersampling.

Figure 3 shows an aliased image that is the zero-filled Fourier reconstruction from the k-space data with the above-introduced undersampling and the averages of 10 and 1000 such images with random sampling. The normalized mean square error (nMSE) is shown at the bottom right corner. These results demonstrate that averages of more images with different sampling patterns can better approximate the ground-truth image.

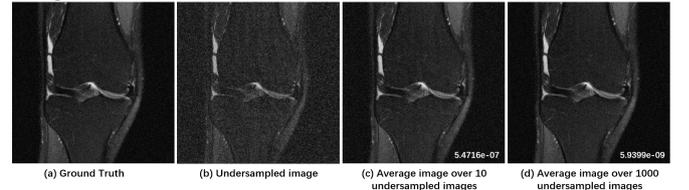

(a) Ground Truth  (b) Undersampled image  (c) Average image over 10 undersampled images  (d) Average image over 1000 undersampled images

Figure 3. Property of the proposed random undersampling pattern. From left to right: (a) ground truth; (b) aliased image from undersampled weighted k-space; (c)-(d) images averaged over 10 and 1000 aliased images from randomly undersampled data, respectively. The nMSE is at the bottom right corner.

## IV. EXPERIMENTAL SETUPS

### A. Dataset

We used the raw single-coil and multi-coil k-space MR data from the knee fastMRI dataset [43] that is available publicly at https://fastmri.org/. The single-coil data in fastMRI is simulated by combining the data from multiple coils into a single coil [44]. In both single and multi-coil cases, we randomly selected 50 sagittal knee MR scans subjects and used them in the training procedure. Out of the total 40 images for each subject, we selected 20 central images that have anatomy, which gave a maximum of 1000 images for training. We further selected another 20 different subjects from the same dataset with 20 central images for each subject, which gave additional 400 images for testing. The dataset with the Turbo Spin Echo sequence has parameters as shown in Table I. All images are complex-valued.

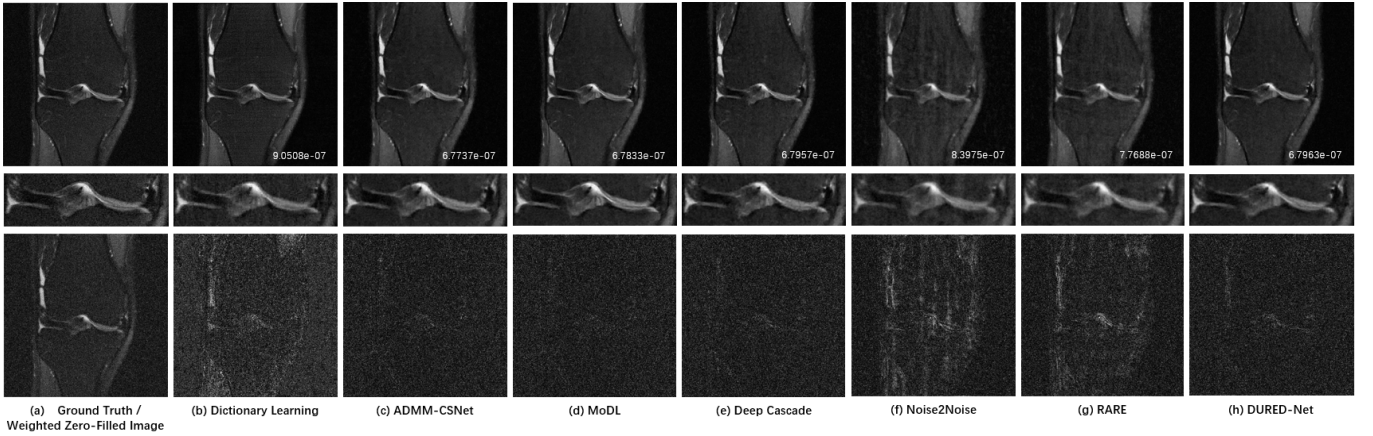

Figure 4. Single-coil reconstruction comparison of the proposed DURED-Net with other networks. From left to right: (a) ground truth (top) and aliased image (bottom), reconstructions with (b) dictionary learning, (c) ADMM-CSNet, (d) MoDL, (e) Deep Cascade, (f) Noise2Noise, (g) RARE, and (h) DURED-Net. From top to bottom: reconstructed images, regions of interest (ROI), and differences (enlarged 4 times) from the ground truth, all on the same scale. The nMSEs are shown at the bottom right corner.

### B. Parameters

The network was trained on a patch size of $320 \times 320$ and mini-batch size of 8 using the Adam algorithm with an initial learning rate of $10^{-3}$. At each epoch, the total number of conjugate gradient iterations in the reconstruction layer $X^n$ to update $\mathbf{x}^n$ was set to 15. The hyperparameters, $\lambda$, and $\beta$ were initialized to $\lambda = 10$ and $\beta = 10$.

To avoid overfitting during the training, data augmentation was used where the images were translated with a random number of pixels vertically and horizontally each time, and the translation was repeated multiple times which equals the number of batched in a training epoch. According to the Fourier transform property, the undersampled k-space data of the translated image can be calculated by multiplying a phase shift term by each k-space data point. Only simple translations were used because other types of data augmentation would change the complex-valued undersampled k-space data in a rather complicated way. Although complex convolution is translation-invariant, the real and imaginary images are separated as two channels in the network and therefore the resulting networking is not translation-invariant.

### C. Quantitative evaluation

We used nMSE and PSNR to evaluate the quality of the reconstructed image quantitatively. Defining $\mathbf{x}_{GT}$ and $\hat{\mathbf{x}}$ to be the ground truth image and the reconstructed image, nMSE is defined as:

$$nMSE(\mathbf{x}_{GT}, \hat{\mathbf{x}}) = \frac{\|\mathbf{x}_{GT} - \hat{\mathbf{x}}\|_2^2}{\|\mathbf{x}_{GT}\|_2^2}. \quad (18)$$

And PSNR is defined as:

$$PSNR(\mathbf{x}_{GT}, \hat{\mathbf{x}}) = 10 \log_{10}\left(\frac{MAX_{\mathbf{x}_{GT}}^2}{MSE(\mathbf{x}_{GT}, \hat{\mathbf{x}})}\right). \quad (19)$$

Table I. Sequence parameters in fastMRI dataset

| Parameter | Value |
|---|---|
| Echo train length | 4 |
| Matrix size | 320×320 |
| In-plane resolution | 0.5mm × 0.5mm |
| Slice thickness | 3mm |
| TR | 2200 ∼ 3000 ms |
| TE | 27 ∼ 34 ms |

## V. RESULTS

### A. Comparison with competing methods

In this section, we compare DURED-Net with different types of deep-learning-based methods as well as the dictionary-learning-based compressed sensing method [4]. The first type is self-supervised learning (trained with the aliased images only) using Noise2Noise [36] and RARE [37]. The other one is the unrolled network with supervised learning using ADMM-CSNet [18], MoDL [21], and the Deep Cascade network [22]. The objective is to show that the proposed DURED-Net is superior to other self-supervise methods, and is close to (but not superior to) the state-of-the-art supervised methods. All supervised methods used the zero-filled image as the network input, and all self-supervised methods used the weighted zero-filled image as the network input according to the requirement in Noise2Noise.

For multi-coil reconstructions, Noise2Noise extended to take 3D data as the input for both training and testing, and the 3D Deep Cascade network was used for both training and testing, where the third dimension represents different channels of coils. No coil sensitivity information was exploited in multi-coil Noise2Noise and Deep Cascade network. In addition, MoDL was used to represent the state-of-the-art supervised unrolled network, which uses a CNN-based regularization prior to the model-based image reconstruction framework. For the proposed multi-coil DURED-Net, the input took the single-coil complex data combined from the multi-coil one as done in [45], and then separated them into a real and imaginary channel. For MoDL, ADMM-CSNet, and DURED-net, the coil sensitivity

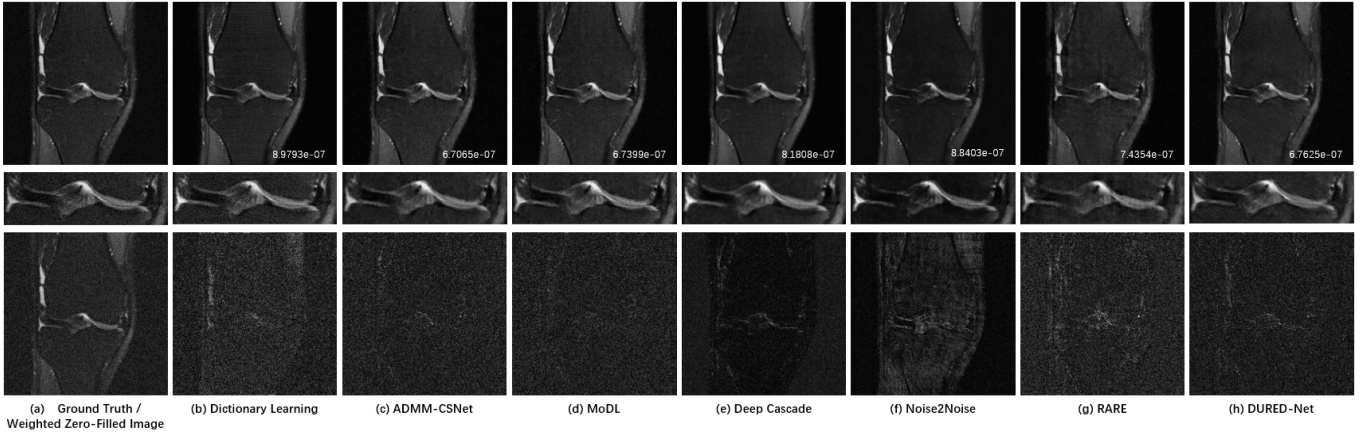

(a) Ground Truth / Weighted Zero-Filled Image  (b) Dictionary Learning  (c) ADMM-CSNet  (d) MoDL  (e) Deep Cascade  (f) Noise2Noise  (g) RARE  (h) DURED-Net information was utilized in the network. DURED-net estimated the sensitivity information using EsPIRiT [46] and then introduced it into the measurement operator **A** when solving for $\mathbf{x}^n$ in Eq.(8).

Figure 5. Multi-coil reconstruction comparison of the proposed DURED-Net with other networks. From left to right: (a) ground truth (top) and zero-filled aliased image (bottom), reconstructions with (b) dictionary learning, (c) ADMM-CSNet, (d) MoDL, (e) Deep Cascade network, (f) Noise2Noise, (g) RARE, and (h) DURED-Net. From top to bottom: reconstructed images, regions of interest (ROI), and differences (enlarged 4 times) from the ground truth, all on the same scale. The nMSEs are shown at the bottom right corner.

TABLE II. Comparison of nMSE (×$10^{-7}$), PSNR (dB), average training time of each epoch, and average testing time with different sampling rates.

| Method | Evaluation | | | | | | | | | | | | Time (s) |
|---|---|---|---|---|---|---|---|---|---|---|---|---|---|
| | Sampling Rate: 10% | | | | Sampling Rate: 20% | | | | Sampling Rate: 50% | | | | Training / epoch \ Testing / slice |
| | Single-Coil | | Multi-Coil | | Single-Coil | | Multi-Coil | | Single-Coil | | Multi-Coil | | |
| | nMSE | PSNR | nMSE | PSNR | nMSE | PSNR | nMSE | PSNR | nMSE | PSNR | nMSE | PSNR | |
| Dictionary Learning | 10.30±5.42 | 24.39±2.91 | 9.53±5.46 | 24.54±3.11 | 7.81±4.07 | 25.24±2.96 | 6.66±3.14 | 27.44±2.44 | 3.46±1.27 | 28.45±1.65 | 2.96±1.27 | 29.45±1.65 | x \ 1380 |
| ADMM-CSNet | 5.67±2.47 | 26.44±1.46 | 5.50±2.46 | 26.58±1.46 | 4.67±2.04 | 27.29±1.49 | 4.74±2.00 | 27.24±1.51 | 2.58±1.12 | 29.87±1.49 | 2.57±1.13 | 29.93±1.51 | 2895 \ 1.24 |
| MoDL | 6.06±2.98 | 26.22±1.65 | 5.95±2.81 | 26.27±1.57 | 4.78±2.14 | 27.19±1.54 | 4.73±2.10 | 27.23±1.53 | 2.61±1.12 | 29.81±1.52 | 2.60±1.10 | 29.82±1.50 | 78.51 \ 0.071 |
| Deep Cascade | 6.48±3.14 | 26.08±1.72 | 7.43±3.20 | 25.47±2.53 | 5.00±2.17 | 27.00±1.58 | 5.48±2.53 | 26.94±2.51 | 2.73±1.15 | 29.61±1.53 | 2.86±1.76 | 29.31±2.38 | 189.53 \ 0.161 |
| Noise2Noise | 9.24±7.13 | 24.69±2.23 | 12.03±9.26 | 23.76±2.76 | 7.52±4.45 | 25.33±1.81 | 8.81±4.67 | 24.90±1.92 | 3.84±1.82 | 28.20±1.83 | 4.22±1.73 | 28.01±1.59 | 38.37 \ 0.005 |
| RARE | 7.50±4.37 | 25.38±2.11 | 9.04±6.01 | 24.75±2.63 | 6.59±3.77 | 26.05±1.55 | 7.27±5.13 | 25.70±2.26 | 3.51±1.55 | 28.58±1.80 | 3.56±1.70 | 28.53±1.74 | x \ 0.045 |
| DURED-Net | 6.60±3.52 | 25.95±1.80 | 6.18±2.25 | 25.91±2.27 | 5.20±2.47 | 26.87±1.68 | 4.86±1.93 | 27.06±1.72 | 2.81±1.16 | 29.53±1.54 | 2.79±0.98 | 29.73±1.51 | 52.57 \ 0.011 |

For supervised learning, the same sampling pattern was used for both training and testing to obtain the best results. All self-supervised learning methods used 1000 training images. For each training image, two different sampling patterns were randomly generated according to Eq. (17) and used to simulate two undersampled scans of the same object. It is worth noting that the performance of the supervised method is significantly worse when the sampling patterns are different in training and testing. Such a requirement for the same sampling patterns may present a challenge in practical MR scans because the actual sampling patterns might vary slightly between scans.

The comparisons were made for both single-coil data, shown in Fig. 4, and multi-coil data, shown in Fig. 5, both from 5x undersampled data using different networks. The DURED-Net reconstruction shows high agreement with the ground truth image, while the images from Noise2Noise and RARE are visibly inferior with loss of sharpness and increased artifacts. The corresponding nMSEs also demonstrate the benefit of DURED-Net quantitatively over other self-supervised methods. We can also see the DURED-Net reconstructions are close to those of the unrolled networks with supervised learning which have the best performance due to the availability of labeled training data and the incorporation of the physics model.

Table II shows the statistical results of different methods, the average training time for one epoch, and the average testing time for one reconstructed slice. It uses 1000 training images and the random sampling mask with an acceleration factor of 2, 5, and 10 for a complex-valued image with a size of 320×320. Compared with the Noise2Noise and RARE, our proposed method produces the best quantitative results with comparable reconstruction time.

### B. Visualization of RED Cost and the Gradient in the proposed method

To analyze the convergence behavior, we visualize the value of the cost functions in Eq. (2) and its gradient field map using the contour graph visualization [47] for the proposed DURED-Net, as well as those for ADMM-RED-based reconstruction with BM3D denoiser and RARE.

Specifically, we start with the reconstructed image $\hat{\mathbf{x}}$ and add random values along two random directions to simulate other reconstructions in the neighborhood of $\hat{\mathbf{x}}$ in a 2D space. Mathematically, it can be represented as:

$$\mathbf{x}_{\alpha,\beta} = \hat{\mathbf{x}} + \alpha \boldsymbol{\varepsilon}_1 + \beta \boldsymbol{\varepsilon}_2 \quad (18)$$

where $\varepsilon_1$ and $\varepsilon_2$ are the 2D random directions following a normal distribution, $\alpha$ and $\beta$ are the randomly chosen values based on the size of the contour graph. For each simulated reconstruction $\mathbf{x}_{\alpha,\beta}$, the corresponding value of the cost function in Eq. (2) is calculated. The same cost values are connected to generate the same contour, and all the contours are shown in the contour graph in Fig. 6 for visualization of the cost function. The original reconstruction ($\alpha = \beta = 0$) is indicated by "+", and the simulated reconstruction with local minimal cost is indicated by "x". In addition, the gradient field of the cost function in Eq. (2) is calculated as

$$\mathcal{J} = \mathbf{A}^T(\mathbf{Ax} - \mathbf{y}) + \lambda(\mathbf{x} - f(\mathbf{x})). \quad (19)$$

The parameter $\lambda$ is empirically optimized in each method. The gradient field direction, as shown as arrows in Fig. 6, indicates the descent direction of the cost function. The center of the gradient field (i.e., the point all the arrows pointing at) represents the global minimum of the cost function, as plotted as "o" in Fig. 6. Note this minimum is different from the local minimum obtained by searching in a 2D space for a contour with the least cost value. If the original reconstruction gives both local and global minimum, then "+", "x", and "o" overlap on the contour graph. Otherwise, the reconstruction algorithm does not generate a local or global optimal solution.

Figure 6 shows the contour graphs and gradient fields for different methods. The first one used the BM3D denoiser in the RED regularization, the second one a pre-trained Noise2Noise denoiser, and the third one our proposed DURED-Net. It can be seen that ADMM-BM3D and RARE reconstructions are away from the local and global minima while the proposed DURED-Net reconstruction agrees with both the local and global optima. This is because only the proposed method updates all components of the cost function using a deep neural network.

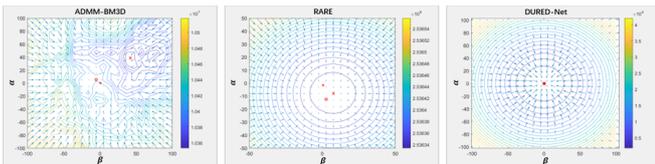

Figure 6. Contours graphs as a function of $(\alpha, \beta)$ for different methods.

### C. Effect of Training Size

In this section, we investigate the influence of training sizes for different deep-learning methods. We compared our method with supervised learning methods, ADMM-CSNet, MoDL, Deep Cascade network, and self-supervised methods, Noise2Noise and RARE. In supervised learning, the same sampling pattern was used for training and testing. Different randomly generated sampling patterns (using Eq. (17)) were used for self-supervised training without ground truth and the subsequent testing. In addition, the training data generated using Eq. (17) were also used to perform self-supervised training for ADMM-CSNet, MoDL, and Deep Cascade networks and the trained models had "-N2N" added at the end.

Figure 7 compares these networks with different training sizes, where the dashed lines with different colors represent the supervised methods and the solid lines represent the self-supervised methods. The plots show the mean PSNRs from 400 reconstructions. We can see all methods benefit from increased training size, but the improvement is the largest for the networks trained without ground truth. This is because the self-supervising learning is based on the Noise2Noise requirement that the mean of aliased images needs to be approximately equal to the clean image. When the training size is relatively small, among all self-supervised methods, the proposed DURED-Net is the closest to the supervised methods. This is because the MR physics model is integrated into the Noise2Noise network in an optimal way for such self-supervised training.

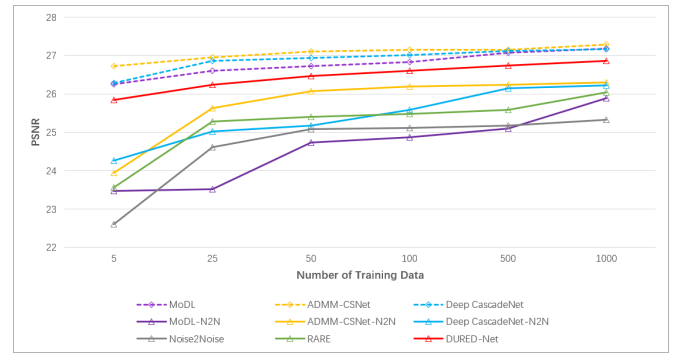

Figure 7. PSNR vs. training size for different networks. An acceleration factor of 5 was used.

### D. Stability Test

We evaluate the stability of different methods in the multi-coil test case by combining two different perturbation tests, as described in [48].

The first test is to add a small but visible perturbation (here we added the letters "CAN U SEE IT" as in [48]) to a test image. The purpose of adding this visible perturbation is to simulate small structural changes caused by pathology that does not exist in the training image, and evaluate if these changes are still as clear in reconstruction as in the original image. As shown in Fig. 8, the added letters in the proposed DURED-Net reconstruction are the sharpest among all self-supervised methods and the sharpness is comparable to MoDL, the supervised ones. It demonstrates that the proposed method is able to recover structures that the network has not seen in the training images.

The second test, named tiny worst-case perturbation, is to add a small and almost-invisible perturbation on top of the first perturbation. The worst-case perturbation can be caused by the imperfection of MR hardware in a practical case. We used the algorithm in [48] to generate the worst-case perturbation. The ground truth image $x$ was perturbed with $\{r_j\}_{j=1}^3$, $|r_1| < |r_2| < |r_3|$, where $|\cdot|$ denotes L1 norm.

Combining those two perturbations simulate practical and challenging cases of data acquisition. We evaluated how much these perturbations deteriorate the reconstructions in Fig. 8. It can be seen that the perturbations corrupt the entire image of both Noise2Noise and RARE reconstructions, especially with the largest perturbation, whereas the proposed DURED-Net is still able to reconstruct the images faithfully even with large perturbations. The above instability tests demonstrate that the proposed DURED-Net is robust to imperfections during the training.

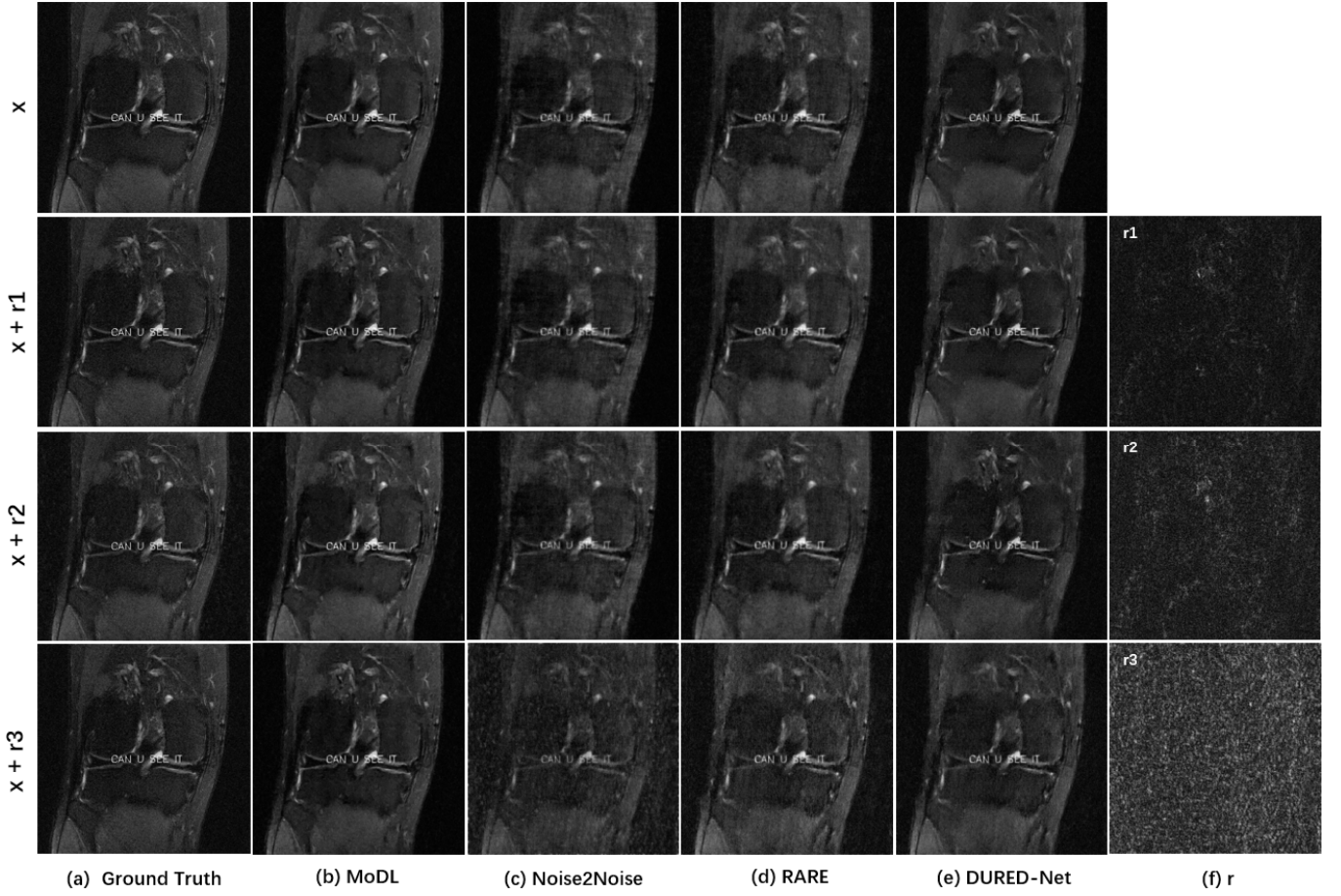

Figure 8. From left to right are the ground truth images, and multi-coil reconstruction of MoDL, Noise2Noise, RARE, and DURED-Net with small visible perturbations and worst-cast perturbations $r_j$ (seventh column), where the worst-case perturbations become larger from top to bottom, $|r_1| < |r_2| < |r_3|$.

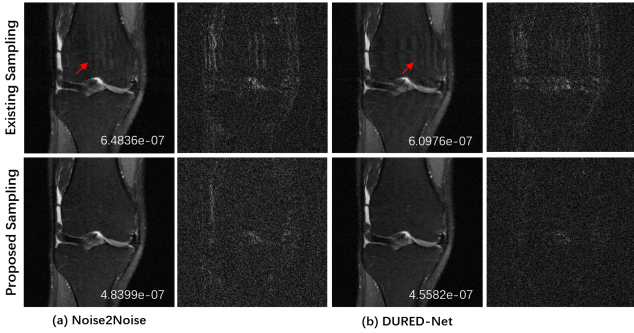

Figure 9. Reconstructions with 1D undersampling (3×) patterns were generated using the method in [36] (top) and the proposed DURED-Net (bottom). (a) The reconstruction of Noise2Noise. (b) The reconstruction of DURED-Net. The difference map was enlarged 4 times. The nMSEs are shown on each image.

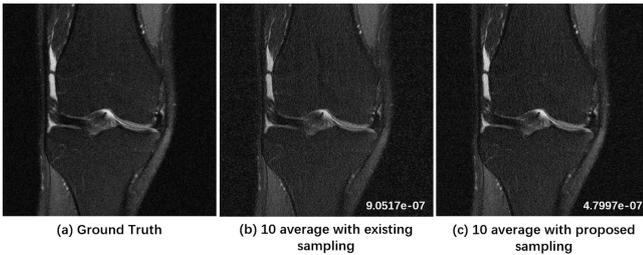

Figure 10. Comparison of the averages of 10 aliased images using two different 1D sampling patterns. The nMSEs are shown.

### E. Reconstruction with 1D sampling patterns

In this section, we evaluate the performance of DURED-Net when 1D random undersampling is used. The single-coil reconstructions of Noise2Noise and DURED-Net with an undersampling factor of 3 are shown in Fig. 9. The top row used the sampling scheme in [36], and the bottom row used the proposed random sampling as described in Eq. (17). A close inspection shows that the existing sampling pattern leads to the artifact as indicated by the red arrow. Whereas the proposed sampling scheme is able to achieve better reconstructions with fewer artifacts. This is because there are two instead of just one parameter in the proposed scheme to adjust the sampling pattern so that the samples are more spread out in k-space for noise-like artifacts. The results also demonstrate that when the same sampling pattern is used, DURED-net is superior to Noise2Noise in reconstructing images from 1D undersampled data, which is typically used in 2D imaging. In addition, we further demonstrate the benefits of the proposed sampling in Fig. 10, where the average of 10 aliased images from randomly undersampled data are compared for the existing and proposed sampling patterns. The result shows that the average using the proposed sampling is closer to the ground truth image, which suggests fewer sampling patterns are needed for training.

### F. Training with fewer sampling patterns

Finally, we investigate the effect of the number of sampling patterns in the training. Because changing to different sampling patterns takes extra time to implement in practice, we evaluate the performance when only a few sampling patterns are used during the training. Figure 11 compares the single-coil reconstructions and their corresponding error maps when 2 and 1000 sampling patterns are used in training. The corresponding statistical evaluation using 400 images is provided in Table III. Both Noise2Noise and DURED-net were used for reconstruction. It can be seen that the Noise2Noise reconstruction loses more details when trained with fewer sampling patterns, whereas the DURED-net reconstruction barely deteriorates. The results indicate that DURED-net only requires two sampling patterns, one for the input and the other for the target during training. And as discussed in Section II.C, the two patterns, which are not necessarily disjoint, can be combined into one pattern such that only a single scan is sufficient.

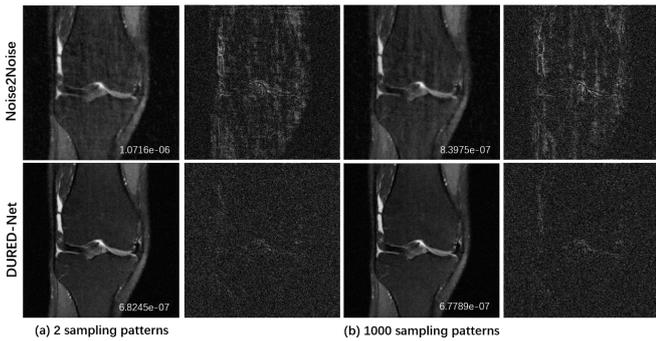

(a) 2 sampling patterns　　　　　　(b) 1000 sampling patterns

Figure 11. Noise2Noise (top) and DURED-Net (bottom) reconstructions when trained with 2 and 1000 different sampling patterns in the target, respectively. The difference map is enlarged 4 times. The nMSEs are shown at the bottom right.

Table III. The statistical evaluation of Noise2Noise and DURED-Net with different numbers of sampling patterns used in training.

| Method | Evaluation | | | |
|---|---|---|---|---|
| | Number of Sampling: 2 | | Number of Sampling: 1000 | |
| | nMSE | PSNR | nMSE | PSNR |
| Noise2Noise | 10.84±7.76 | 24.00±2.26 | 7.52±4.45 | 25.33±1.81 |
| DURED-Net | 5.32±2.41 | 26.56±1.67 | 5.20±2.47 | 26.87±1.68 |

## VI. CONCLUSION

In this paper, we proposed a novel self-supervised learning method DURED-Net for MRI reconstruction from undersampled k-space data. The method effectively integrates Noise2Noise, a self-supervised learning method, with RED, a plug-and-play method, by unrolling the underlying optimization algorithm. We also proposed an under-sampling scheme to improve the performance of the reconstruction method. Results demonstrate the proposed DURED-Net can alleviate the overfitting problem in state-of-the-art methods. The method is robust to small disturbances that are absent during training. Since different sampling patterns are used in training, the method has the benefit that reconstruction can use sampling patterns that are different from those used during training, avoiding artifacts due to sampling pattern mismatch in supervised training. The proposed method will be useful in imaging fast-moving objects where fully sampled labels are challenging to acquire.